\RequirePackage{fix-cm}
\documentclass[]{svjour3}
\smartqed  
\usepackage{setspace}
\usepackage{graphicx}

\usepackage{amssymb}

\usepackage{rotating}

\usepackage{natbib}

\newcommand{\bfg}[1]{\mbox{\boldmath $#1$\unboldmath}}

\newcommand{\fraca}[2]{\displaystyle\frac{#1}{#2}}

\def \R {{\rm I\kern -2.2pt R\hskip 1pt}}

\newtheorem{algorithm1}{Algorithm}

\def\boxforqed{\rule{0.5em}{1.5ex}}
\def\qed{\ifmmode\squareforqed\else{\unskip\nobreak\hfil
        \penalty50\hskip1em\null\nobreak\hfil\boxforqed
         \parfillskip=0pt\finalhyphendemerits=0\endgraf}\fi}

\journalname{}
\begin{document}

\title{Mixed extreme wave climate model for reanalysis data bases
}
\subtitle{}


\author{R.~M\'{\i}nguez \and A~.~Tom\'as \and F.~J.~M\'endez \and R.~Medina
}


\institute{All authors \at
              Environmental Hydraulics Institute ``IH Cantabria'', Universidad de Cantabria,
Cantabria Campus Internacional, Spain\\
              Tel.: +34-942-201852\\
              Fax: +34-942-201860 \\
              \email{roberto.minguez@unican.es}           
}

\date{Received: date / Accepted: date}

\maketitle

\begin{abstract}
Hindcast or Wave Reanalysis Data Bases (WRDB) constitute a powerful source with respect to instrumental records
for the design of offshore and coastal structures, since they offer important
advantages for the statistical characterization of wave climate variables, such as
continuous long time records of significant wave heights,
mean and peak periods, etc. However, reanalysis data is less accurate than instrumental records,
which makes extreme data analysis derived from WRDB prone to under predict design
 return period values. This paper proposes
a Mixed Extreme Value (MEV) model to deal with maxima that takes full advantage of both i) hindcast or wave reanalysis,
 and ii) instrumental records, reducing the uncertainty on its predictions. The resulting mixed
 model
 merges consistently the information given by
  both kind of data sets, and it can be applied to any extreme value analysis
  distribution, such as GEV or Pareto-Poisson.
The methodology is illustrated using both synthetically generated and real data,
 the latter taken
 from a given location in the Northern Spanish coast.
\end{abstract}

\section{Introduction}\label{s1}
Extreme value analysis is of paramount importance for the design process of coastal and offshore
 structures. The objective of the design is to verify that the structure satisfies the project requirements
 during its lifetime in terms of acceptable failure rates and costs. One of these project requirements is
 the ultimate limit state design, i.e.
the structure must withstand the maximum stresses which are expected to occur during the lifetime.
The appropriate definition of this ultimate limit state design relies on the correct evaluation of the wave
climate producing the worst case scenario, i.e. on extreme wave climate analysis.

Over the last decade, in an attempt to improve the knowledge about wave climate, there has been an outstanding development of wave reanalysis models. These models
 allow a detailed description of wave climate in locations where long-term buoy records
do not exist. This fact has raised the attention of scientists and engineers, which have tried to use them
for design purposes. However, several authors (\cite{CairesS:05,CavaleriS:06,MinguezETML:11,MinguezRLM:11}) have pointed out discrepancies when comparing reanalysis versus instrumental data. The reasons are multiple:
numerical models are simplifications of reality, discrete spatio-temporal resolutions,
temporal resolutions too coarse (6 hours) to include high-frequency energy,
complexity of the orography in certain regions, bathymetric deficiencies.
These discrepancies are specially relevant in shallow waters, and during the occurrence of hurricanes and typhoons, where the model resolutions are not enough to reproduce the physics.

In order to reduce these discrepancies, several authors have attempted to combine
reanalysis versus instrumental observations, taking full advantage of the goodness of both kind of information. For example, \cite{CairesS:05} establish a nonparametric correction based on ``analogs'', \cite{CavaleriS:06} calibrate decadal time series over the Mediterranean Sea using  buoys and satellites, \cite{TomasML:08} propose a spatial calibration procedure
based on empirical ortogonal functions and a non-linear transformation of the spatial-time modes,
\cite{MinguezETML:11} present a nonlinear regression model for directional calibration.
However, these methods perform appropriately for most of the range of the probability density function
of the reanalysis variables but for extremes. In fact, \cite{MinguezRLM:11} demonstrates the importance
 of removing these extreme events from the calibration process, since
in certain occasions they may distort the calibration procedure and still not solving the extreme event discrepancies.
In addition, they introduce several regression models for automatic detection of these events,
before removing them from the calibration process.

The statistical theory of extreme values \citep{Castillo:88,Coles:01,KatzPN:02,CastilloHBS:05} provides the
mathematical framework for modeling the tail distribution, i.e. the extreme values,
when maximum datasets are available. These models allow us to obtain useful information, such as
return period values for certain variables. Several models and applications
have been used in different climate studies to model
block extremes, typically annual maxima or minima, both in observed and
simulated data \citep{KharinZZ:05,GoubanovaL:06,
KioutsioukisMZ:10,NikulinKHSU:11}.
In addition, recent advances in extreme value theory allow introducing time-dependent
variations in the extreme value models. In this kind of approach, parameters are
replaced by different functions dependent of time \citep{Coles:01}.
In a simple setting, the parameters can include a
trend term varying linearly with time \citep{Cooley:09} or a forcing term
varying with some external climatic indices, such as the Southern Oscillation Index
or the North Atlantic Oscillation (NAO). There are also studies combining both
approaches \citep{MendezMLL:07}. The most complex approaches consider harmonic functions
reflecting the seasonality of the occurrence of maxima.  For instance,
\citet{MenendezMIL:09,IzaguirreMMLL:10,MinguezMIML:10} developed a time-dependent model
based on the GEV distribution that accounts for the seasonality and interannual
variability of extreme wave height. 
 A similar approach has been considered
by \citet{RustMO:09} to model extreme precipitation in the UK on
a seasonal basis.
\cite{GaliatsatouP:11} present a statistical model for extreme value analysis
considering seasonality. They use a non-stationary point process approach
 estimated through the wavelet transform.
\cite{Vanem:11} presents a literature survey on time-dependent statistical
modelling of extreme waves and sea states.

The main problem, from the engineering design perspective, is that neither
i) none of the previous
calibration/correction approaches, nor ii) the extreme value analysis models proposed in the literature,
provide the answer about how to deal with extreme events appropriately in the case of dealing with
reanalysis and instrumental data.
 The aim of this paper is to fill this niche
 by presenting a general method to deal with extremes that takes full advantage of both i) hindcast or
 wave reanalysis, and ii) instrumental records.
The resulting model merges consistently the information given by
  both kind of data sets, reducing the uncertainty on its predictions. In addition, it can
   be applied to any extreme value analysis distribution, such as GEV or Pareto-Poisson.

The paper is organized as follows.
Section~\ref{s3} presents the proposed Mixed Extreme Value model, while in Section~\ref{s4}
several diagnostic tests are given to check the appropriateness of the model hypothesis.
In Section~\ref{s5}, the functioning of the method is illustrated through
two different simulation experiments.
In contrast, Section~\ref{s6} shows the performance on real data from a given location in the North
of Spain.
Finally, in Section \ref{s7} relevant conclusions are drawn.

\section{Mixed Extreme Value Analysis Model}\label{s3}
Extreme value analysis concerns the knowledge about the occurrence of extreme events and their frequency,
and a careful analysis requires the availability of data on such extremes \citep{CastilloCM:08}.
The larger the size of the data record, the more accurate the statistical model for those extremes
will be, which would lead to better predictions with lower uncertainties. \cite{Vanem:11} points out
the importance of available wave data in order to develop adequate
probabilistic models, and although buoy measurements are generally regarded as most reliable,
alternatives exist in satellite data and in reanalysis data obtained from wave models forced by
various meteorological parameters. However, discrepancies between numerical and instrumental
records must be accounted for within the analysis.

\begin{figure*}
\begin{center}
\includegraphics*[width=0.90\textwidth]{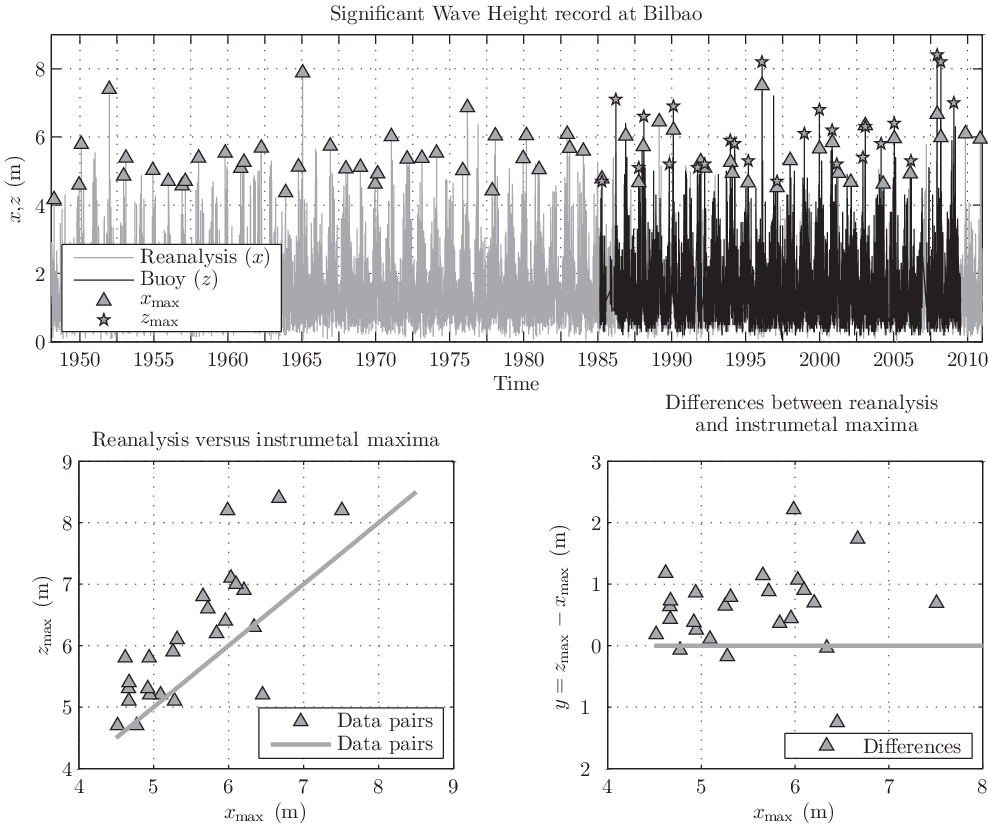}
\caption{\label{f1} Instrumental and reanalysis significant wave height records for Bilbao buoy location.}
\end{center}
\end{figure*}

Figure~\ref{f1} shows the significant wave height instrumental (black line, taken from
Puertos del Estado (Spain) buoys database) and reanalysis (gray line) records
 for Bilbao (Spain) buoy location, and their corresponding annual maxima (triangle and star dots, respectively).
Reanalysis data is taken from Downscaled Ocean Waves (DOW) database, which constitutes
a numerical wave database propagated to the Spanish coastal areas from the Environmental
Hydraulics Institute (Spain). The DOW database is a hybrid downscaling \citep{CamusMM:11}
 from the GOW hindcast database (Global Ocean Waves, \cite{RegueroMMML:11}).

The most reliable and accurate maxima corresponds to instrumental data, in fact, for years where both maxima are available,
the differences between them are shown in the panels below. These panels represent the reanalysis versus instrumental
maxima, and reanalysis maxima versus the differences between them, respectively.
If we make extreme value (EV) analysis using i) instrumental data and ii) reanalysis data, respectively. The instrumental
EV model would be more reliable with respect to expected return periods, but the uncertainty would increase
with respect to that of the reanalysis
EV model, because the latter uses more data. The model presented in this paper allows using both instrumental and reanalysis
data, which results in a more robust estimation of return period values decreasing the uncertainty.

Let consider the maximum annual records from reanalysis (${\bfg x}_{\rm max}$) and instrumental (${\bfg z}_{\rm max}$),
with lengths $n_x$ and $n_z$, respectively. Note that we assume that $n_z<<n_x$. For years where both maxima
are available, we get the vector ${\bfg y}_{\rm max}$ as the differences between instrumental and reanalysis maxima.
The proposed mixed model relies on the following assumptions:
\begin{enumerate}
  \item \label{Hypo1}The annual maximum reanalysis random variable $X$ follows distribution with probability
density and cumulative distribution functions $f_{X}(x,{\bfg \theta}_X)$ and $F_{X}(x,{\bfg \theta}_X)$,
respectively. The distribution function may correspond to any kind of distribution for maxima, such as, GEV,
 Pareto-Poisson, Gumbel, etc.
 \item \label{Hypo2}The random variable $Y$ corresponding to the difference between instrumental and reanalysis data
  conditioned to the reanalysis maximum data ($X$)
 follows a normal distribution, i.e. $f_{Y|X}(y)\sim N\left(\mu_{Y|X},\sigma_{Y|X}^2\right)$. Note that
 $\mu_{Y|X}$ and $\sigma_{Y|X}$ correspond to the conditional mean and standard deviation parameters, which can be obtained
 using an heteroscedastic regression model.
\end{enumerate}

The annual maximum instrumental random variable is equal to $Z=X+Y$,
and their corresponding cumulative distribution function is equal to:
\begin{equation}\label{FZcum1}
    F_{Z}(z)=\mbox{Prob}(Z\le z)=\int\limits_{x+y\le z}f_{X,Y}(x,y)dydx,
\end{equation}
where $f_{X,Y}(x,y)$ is the joint probability density function of the random variables $X$ and $Y$.
Considering assumptions~\ref{Hypo1} and \ref{Hypo2}, expression~(\ref{FZcum1}) becomes:
\begin{equation}\label{FZcum2}
    F_{Z}(z)=\int\limits_{-\infty}^\infty f_{X}(x,{\bfg \theta}_X) \left[\int\limits_{-\infty}^{z-x}  f_{Y|X}(y)dy\right]dx,
\end{equation}
and since the distribution of $Y$ conditioned to $X$ is normally distributed, expression~(\ref{FZcum2})
results in:
\begin{equation}\label{FZcum3}
    F_{Z}(z)=\int\limits_{-\infty}^\infty f_{X}(x,{\bfg \theta}_X) \Phi\left[\fraca{z-x-\mu_{Y|X}}{\sigma_{Y|X}}\right]dx,
\end{equation}
where $\Phi(\cdot)$ is the cumulative distribution of the standard normal random variable.

The corresponding probability density function is obtained deriving (\ref{FZcum3}) with respect to $z$:
\begin{equation}\label{fZcum}
    f_{Z}(z)=\int\limits_{-\infty}^\infty f_{X}(x,{\bfg \theta}_X) \phi\left[\fraca{z-x-\mu_{Y|X}}{\sigma_{Y|X}}\right]\fraca{1}{\sigma_{Y|X}}dx,
\end{equation}
where $\phi(\cdot)$ is the probability density function of the standard normal random variable. Note that
the integration limits range from $-\infty$ to $\infty$, however, these limits may change depending on the
type of $X$ probability density function.

Both PDF~(\ref{fZcum}) and CDF~(\ref{FZcum3}) require to solve an integral over a varying domain.
This task can be efficiently achieved using numerical quadrature methods. Numerical tests performed, using different methods,
indicate that the adaptive Gauss-Kronrod quadrature method \citep{Shampine:08} is the most appropriate, since it
supports infinite intervals and can handle moderate singularities at the endpoints.

The corresponding quantile $z_{q}$ for a given probability $q$ is obtained by solving the following implicit equation:
\begin{equation}\label{quan}
F_{Z}(z_{q})=q,
\end{equation}
which can be transformed into the problem of finding the root of the function $g(z_q)=q-F_{Z}(z_{q})$.
Numerical tests indicate that the algorithm proposed by \cite{ForsytheMM:76}, which uses
 a combination of bisection, secant, and inverse quadratic interpolation methods,
 is robust and efficient.

 An important feature of any EV model corresponds to the calculation of the
 confidence intervals on both model parameters and estimates, i.e. uncertainty. Since the proposed mixed model depends on two
 independent distributions, previously to analyze confidence intervals related to $Z$, we will briefly describe
 how to deal with $f_X(x)$ and $f_{Y|X}(y)$ distributions, and their confidence intervals. Finally,
 estimated parameters  and quantile confidence intervals for the proposed model are given in
Appendixes~\ref{apenA} and \ref{apenB}, respectively.

\subsection{Reanalysis annual maxima distribution ($f_X(x)$)}
One of the advantages of the proposed mixed model is that annual maxima can be analyzed using any extreme value
analysis distribution. In this paper we only provide expressions and examples for the Generalized Extreme Value (GEV) distribution
and the Pareto-Poisson model \citep{Castillo:88}.

\subsubsection{GEV Distribution}\label{GEV}
 Within this approach, annual maxima of
successive years are assumed to be i) independent random variables
and ii) identically distributed.
Annual maximum $X$ of the climate variable
follows a GEV distribution with time-dependent
location parameter $\mu$, scale parameter $\psi$, and
shape parameter $\xi$, with a cumulative distribution function (CDF) given by:
\begin{equation}\label{cdfGEV}
F_X(x;\mu ,\psi,\xi)=\left\{ \begin{array}{l}\exp \left\{
-\left[ 1+\xi\left( \fraca{x-\mu}{\psi } \right)
\right]_+^{-\fraca{1}{\xi}} \right\};  \xi \ne
0,\nonumber\\  \\
\exp\left\{-\exp\left[-\left(\fraca{x-\mu}{\psi}\right)\right]\right\};
   \xi=0
,
\end{array}\right.
\end{equation}
where $[a]_+=\max(0,a)$, and the support is $x \le \mu -
\psi/\xi$, if $\xi < 0$, or $x \ge \mu -
\psi/\xi$, if $\xi > 0$. The GEV family includes three distributions corresponding
to the different types of tail behavior: Gumbel ($\xi=0$) with a light tail decaying
exponentially; Fr\'echet distribution ($\xi > 0$) with a heavy tail
decaying polinomially; and Weibull ($\xi < 0$) with a bounded
tail.

From (\ref{cdfGEV}), the corresponding quantiles $x_q$, where $q$ is the corresponding probability,
 can be straightforwardly calculated.

Model parameters ${\bfg \theta}_X=(\mu, \psi,\xi)^T$ may be estimated using the method
of maximum likelihood.

\subsubsection{Pareto-Poisson Distribution}\label{ParPoi}
This model combines the Generalized Pareto Distribution (GPD) for studying exceedances over a threshold $u$, and the
Poisson distribution for occurrence of exceedances. It is based on the following assumptions:
\begin{enumerate}
  \item The number of exceedances over the level $u$ during the year has a Poisson distribution with parameter $\lambda$.
  \item Those exceedances follow the GPD distribution.
\end{enumerate}

Under these hypothesis, the cumulative probability distribution (CDF) of the annual maximum can be expressed as:
\begin{equation}\label{cdfParPoi}
F_X(x;\lambda ,\psi,\xi)=\left\{ \begin{array}{l}\exp \left\{
-\lambda\left[ 1+\xi\left( \fraca{x-u}{\psi } \right)
\right]_+^{-\fraca{1}{\xi}} \right\};  \xi \ne
0,\nonumber\\  \\
\exp\left\{-\lambda\exp\left[-\left(\fraca{x-u}{\psi}\right)\right]\right\};
   \xi=0
,
\end{array}\right.
\end{equation}
where $[a]_+=\max(0,a)$, and the support is $x>u$, $x \le
\psi/|\xi|$ if $\xi < 0$, or $x \le \infty$ if $\xi > 0$.
The Pareto-Poisson family includes, as the Pareto family, three distributions corresponding
to the different types of tail behavior: Exponential ($\xi=0$); traditional Pareto tail ($\xi > 0$),
and an analogous Weibull distribution ($\xi < 0$) with a bounded
tail.

Analogously to the GEV model, quantiles and estimates can be obtained from (\ref{cdfParPoi})
and the method of maximum likelihood, respectively.

\subsection{Heteroscedastic regression model ($f_{Y|X}(y)$)}
Consider the standard nonlinear regression model
\begin{equation}\label{e.lgm}
{\bfg y} = f_{\mu}({\bfg x},{\bfg \beta}_\mu) +  {\bfg \varepsilon},
\end{equation}
where ${\bfg y}=(y_1,y_2,\ldots,y_{n_y})^T$ is the $n_y\times 1$ response variable vector
 associated with the differences between instrumental and reanalysis maxima,
 ${\bfg x}$ is a $n_y\times 1$ vector of predictor variables related to
  annual reanalysis maxima,
the function $f_\mu$ is known and nonlinear in the parameter
vector ${\bfg \beta}_\mu$, and $\varepsilon_i;i=1,\ldots,n_y$ are jointly normally distributed
$\bfg \varepsilon\sim N({\bf 0},\sigma^2{\bfg V})$ errors,
where $\sigma^2{\bfg V}$ is a positive definite variance-covariance matrix.

In the standard Nonlinear Least Square (NLS) method,
the parameter estimation problem can be stated as:
\begin{equation}\label{e.nlLS2}
\mathop{{\mbox{Minimize}}}\limits_{{{\bfg \beta}}}\; \bfg \varepsilon^T (\sigma^2{\bfg V})^{-1}  \bfg \varepsilon.
\end{equation}

However, for the kind of data considered in this regression model, a simple scatter plot of differences ($y$)
 versus reanalysis data ($x$) allows
observing how the variance of the regression model may change over the regression
function  \citep{MinguezRLM:11}. Consequently, we consider
a nonlinear heteroscedastic regression model in which the standard deviation $\sigma_i$ of the $i$th error
is a function of the predictor variable ($x_i$):
\begin{equation}\label{sigeq}
    \sigma_i = f_\sigma(x_i;{\bfg \beta}_\sigma),
\end{equation}
where ${\bfg \beta}_\sigma$ is a vector of coefficients or parameters.
The parameter vector ${\bfg \beta}=[{\bfg \beta}_\mu; {\bfg \beta}_\sigma]$, of size $n_p \times 1$,
can be estimated maximizing the log-likelihood function:
\begin{equation}\label{Rmaxlikel}
    \ell({\bfg \beta};{\bfg x},{\bfg y})=-\sum_{i=1}^{n_y}\log\left(f_\sigma(x_i;{\bfg \beta}_\sigma)
    \right)-\fraca{1}{2}\sum_{i=1}^{n_y}\left(\fraca{y_i - f_\mu(x_i; {\bfg \beta}_\mu)}{f_\sigma(x_i;{\bfg \beta}_\sigma)}\right)^2.
\end{equation}

The advantage of defining the regression model in a general way is that it allows using different parameterizations
for the mean and standard deviation, for instance, possible models are:
\begin{eqnarray}
f_{\mu}(x_i,{\bfg \beta}_\mu)=\beta_0 + x_i {\beta_1};&& f_{\sigma}(x_i,{\bfg \beta}_\sigma) = \beta_2 + x_i {\beta_3},\label{model1}\\
  f_{\mu}(x_i,{\bfg \beta}_\mu)=\beta_0 x_i^{\beta_1};&& f_{\sigma}(x_i,{\bfg \beta}_\sigma) = \beta_2 x_i^{\beta_3},\label{model2}
\end{eqnarray}
but any other expression could be used instead. Note that both models (\ref{model1}) and (\ref{model2}) include the classical
homoscedastic
linear regression model provided that $\beta_3=0$.

\section{Hypothesis testing}\label{s4}
The mixed model proposed in this paper is based on several assumptions.
 For this reason, once the parameter estimation processes for both i) the EV model over $x$ and ii) the
 regression model over $y$ conditional on $x$, are finished, it is very important to make and run
different diagnostic plots and statistical hypothesis tests to check whether the selected
distributions were appropriate or not.

\subsection{EV model for $X$}
Related to the EV model fitted to the reanalysis data, we use the following diagnostic plots and tests:
\begin{itemize}
  \item Probability-probability (PP) and Quantile-quantile (QQ) plots. Points over the diagonal are indicative of
  a good quality fit.

  \item A one-sample Kolmogorov-Smirnov test (\cite{Massey:97}). This test compares, for a given significance
  level $\alpha$, the transformed values
  ${\bfg x}^{{\rm N}}$ using transformation ${\bfg x}^{{\rm N}} = \Phi^{-1}\left[F_{X^{}}({\bfg x}_{\rm max})\right]$ with respect
   to a standard
  normal distribution. The null hypothesis is that the transformed sample follows a standard normal distribution.

  \item Sample autocorrelation and partial autocorrelation functions related to the
    transformed sample ${\bfg x}^{{\rm N}}$. These functions help checking the independence assumption between maxima,
    whose values should be within the confidence bounds.

    \item To further explore the independence hypothesis, the Ljung-Box lack-of-fit hypothesis test
    \citep{BrockwellD:91} for model
     misidentification is applied. This test indicates the acceptance or not of the null
     hypothesis that the model fit is adequate (no serial correlation at the corresponding
      element of Lags).

 \end{itemize}

 \subsection{Regression model for $Y|X$}
Related to the heteroscedastic conditional regression model, the basic assumption for this model to be considered appropriate,
is that studentized residuals are independent and normally distributed. Studentized residuals are computed as:
\begin{equation}\label{normres}
    \hat\varepsilon_i^{\rm N}
    = \frac{\hat \varepsilon_i}{\sqrt{\Omega_{i,i}}}
    =\frac{y_i-f_\mu(x_i;\;\hat{\bfg \beta}_\mu)}{\sqrt{\Omega_{i,i}}} \quad  i=1,\dots,n_y,
\end{equation}
where $\Omega_{i,i}$ is the $i$th diagonal element of the residual variance-covariance matrix $\bfg \Omega$.
Details about the derivation of this matrix can be found in \cite{MinguezRLM:11}.

According to that, we use the following diagnostic plots and tests:
\begin{itemize}
  \item A one-sample Kolmogorov-Smirnov test (\cite{Massey:97}) to to check, for a given significance
  level $\alpha$, that studentized residuals follow a standard
  normal distribution. The null hypothesis is that they do.

  \item Sample autocorrelation and partial autocorrelation functions related to studentized residuals.
   These functions helps checking the independence assumption,
    whose values should be within the confidence bounds.

    \item To further explore the independence hypothesis, the Ljung-Box lack-of-fit hypothesis test
    \citep{BrockwellD:91} for model
     misidentification is applied. This test indicates the acceptance or not of the null
     hypothesis that the model fit is adequate (no serial correlation at the corresponding
      element of Lags).

 \end{itemize}

Additional or alternative tests for those selected above could be applied. Note that in case any
of those tests allow rejecting the null hypothesis with a given significance level, the probability
distribution assumptions must be revisited before being acceptable
for return period predictions.

\section{Simulation Case Study}\label{s5}
Previous to the application of the proposed method to a realistic case, we will perform a simulation study to check
whether the method provide consistent results when the data follows the required assumptions. We
consider two simulated samples with the following characteristics:
\begin{description}
  \item[Case 1:] The reanalysis simulated sample follows a GEV distribution with parameters ${\bfg \theta}_X^{\rm true}=(10,\exp(0.5),-0.15)^T$, and the
  heteroscedastic model correspond to that given in (\ref{model1}) with parameters ${\bfg \beta}^{\rm true} = (-0.5, 0.7, -0.3, 0.1)^T$.
  The samples ${\bfg x}^{\rm max}_1$ and ${\bfg y}_1$ have $n = 1000$ records each, which corresponds to
  a simulated period of 1000 years.

  \item[Case 2:] The reanalysis simulated sample follows a Pareto-Poisson distribution with parameters
  ${\bfg \theta}_X^{\rm true}=(25,\exp(-0.13),-0.05)^T$, where
  the first parameter corresponds to the expected number of exceedances per year, i.e. $\lambda$, and the remainder to the GPD
  distribution parameters. The threshold is equal to u = 2.5. Regarding the
  heteroscedastic model, it also corresponds to that given in (\ref{model1}) with parameters
  ${\bfg \beta}^{\rm true} = (0.16, 0.04, 0.3, 0.06)^T$.
  In order to get 1000 years of maximum data, we sample $n = 1000\times\lambda^{\rm true}=25000$ records of exceedances
  using the given GPD distribution, which constitutes the sample ${\bfg x}_2$. The maximum for each year constitutes
  the sample ${\bfg x}^{\rm max}_2$ and the associated difference with respect
   to instrumental data corresponds to ${\bfg y}_2$.
\end{description}

With both sample records, we proceed to perform the following steps:
\begin{description}
  \item[Case 1:] Using the first simulated sample set $({\bfg x}^{\rm max}_1,{\bfg y}_1)$:
      \begin{enumerate}
        \item We fit the reanalysis sample to the GEV by maximizing the log-likelihood function, getting the following
        parameter estimates and 95\% confidence bounds:
         \begin{equation}\label{fitGEV1}
             \begin{array}{rcl}
               \hat \mu_1 & = & 9.9695\;(9.9659,9.9731) \\
               \hat \psi_1 & = & 0.4952\;(0.4936,0.4967) \\
               \hat \xi_1 & = & -0.1477\;(-0.1490,0.1465), \\
              \end{array}
         \end{equation}
which, as expected, are very close to the true values used for simulation purposes.

    \item We fit the sample $({\bfg x}^{\rm max}_1,{\bfg y}_1)$ to the conditional regression model
     by maximizing the log-likelihood function (\ref{Rmaxlikel}), getting the following
        parameter estimates and 95\% confidence bounds:
         \begin{equation}\label{fitHET1}
             \begin{array}{rcl}
               \hat \beta_1 & = & -0.5541\;(-0.8022,-0.3059) \\
               \hat \beta_2 & = & 0.7049\;(0.6801,0.7296) \\
               \hat \beta_3 & = & -0.2836\;(-0.4621,-0.1051) \\
               \hat \beta_41 & = & 0.0949\;(0.0771,0.1128) \\
              \end{array}
         \end{equation}
which, as expected, are very close to the true values used for simulation purposes.
     \item Additionally, we fit the instrumental sample, i.e.
     ${\bfg z}^{\rm max}_1 = {\bfg x}^{\rm max}_1+{\bfg y}_1$ to a GEV distribution, getting the following
        parameter estimates and 95\% confidence bounds:
         \begin{equation}\label{fitGEV2}
             \begin{array}{rcl}
               \hat \mu_1 & = & 16.3913 \;(16.3850,16.3976) \\
               \hat \psi_1 & = & 1.0494 \;(1.0479,1.0510) \\
               \hat \xi_1 & = & -0.1357\;(-0.1369 ,-0.1345). \\
              \end{array}
         \end{equation}

    \item Using the three fitted models, the quantiles and their 95\% confidence intervals related to i) the reanalysis ($X$) data fit,
     ii) instrumental ($Z$) data fit using the proposed model, and iii) instrumental ($Z$) data fit using the GEV fit, are calculated.
      \end{enumerate}

      Results for this case are shown in Figure~\ref{QuantilesGEV}, where the following observations are pertinent:
      \begin{enumerate}
        \item The three models provide very good fits with respect to simulated data, which was an expected result due to
        the length of the data samples.
        \item The proposed mixed model provides almost undistinguishable expected quantiles with respect to the GEV fit to maxima
        $z$-values. This proves the consistency of the proposed model.
        \item In terms of confidence intervals, the proposed model provides slightly better results. Note that our proposed model
        confidence intervals are always between the GEV confidence intervals.
      \end{enumerate}

\begin{figure}
\begin{center}
\includegraphics*[width=0.95\textwidth]{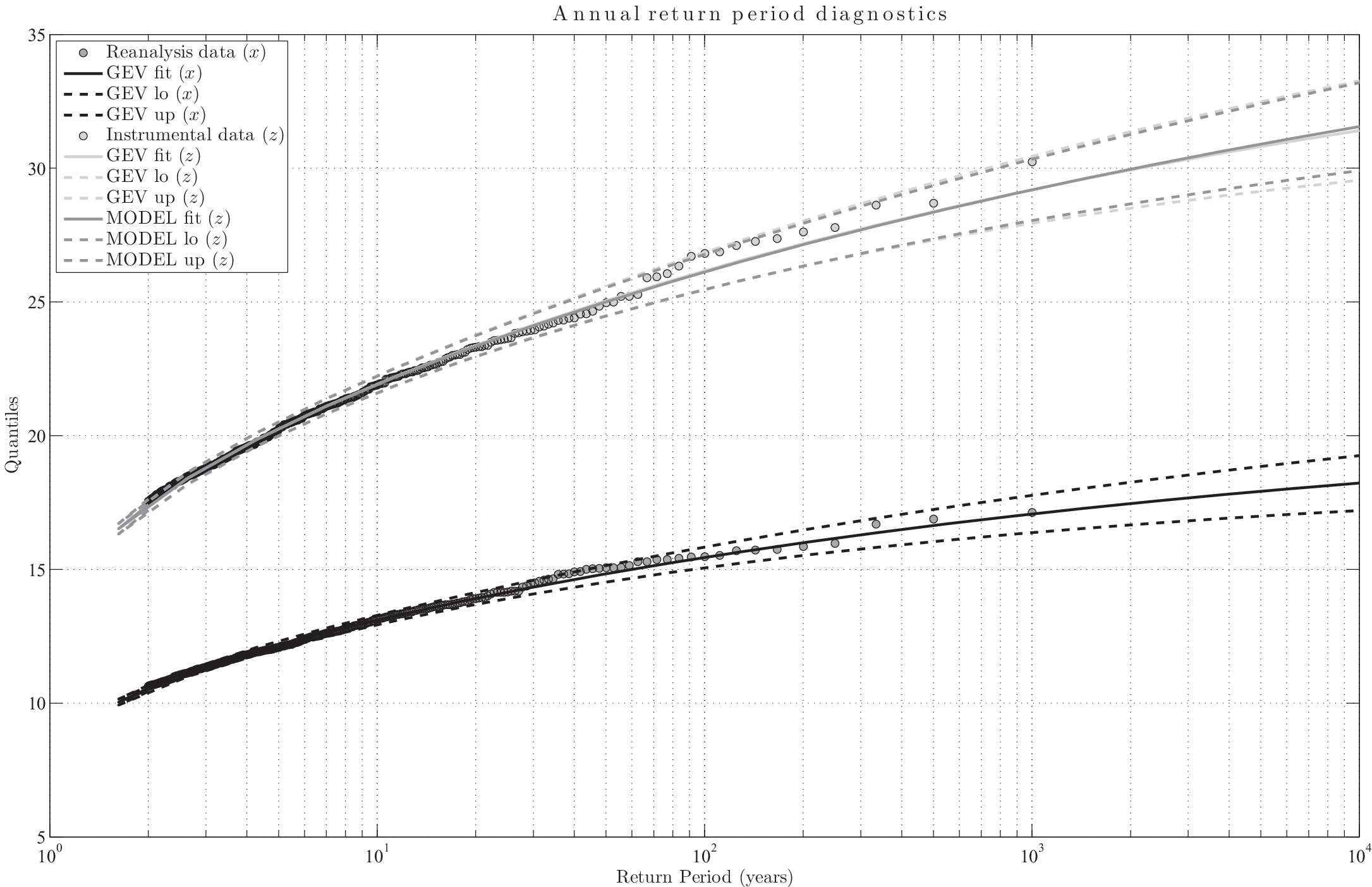}
\caption{\label{QuantilesGEV}Simulated data, quantiles and their 95\% confidence intervals related to i) the reanalysis ($X$) data fit,
     ii) instrumental ($Z$) data fit using the proposed model, and iii) instrumental ($Z$) data fit using the GEV fit.}
\end{center}
\end{figure}

  \item[Case 2:] Analogously to the previous case, using the second simulated sample set $({\bfg x}_2,{\bfg x}^{\rm max}_2,{\bfg y}_2)$:
      \begin{enumerate}
        \item We fit the reanalysis sample ${\bfg x}_2$ to the Pareto-Poisson distribution
        by maximizing the log-likelihood function, getting the following
        parameter estimates and 95\% confidence bounds:
         \begin{equation}\label{fitPOT1}
             \begin{array}{rcl}
               \hat \lambda_2 & = & 25.0000\;(24.9373,25.0627) \\
               \hat \psi_2 & = & -0.1438\;(-0.1444,-0.1433) \\
               \hat \xi_2 & = & -0.0381\;(-0.0381,-0.0374), \\
              \end{array}
         \end{equation}
which, as expected, are very close to the true values used for simulation purposes.

    \item We fit the sample $({\bfg x}^{\rm max}_2,{\bfg y}_2)$ to the conditional regression model
     by maximizing the log-likelihood function (\ref{Rmaxlikel}), getting the following
        parameter estimates and 95\% confidence bounds:
         \begin{equation}\label{fitHET2}
             \begin{array}{rcl}
               \hat \beta_1 & = & 0.1334\;(-0.1257,0.3926)\\
               \hat \beta_2 & = & 0.0481\;(0.0006,0.0957) \\
               \hat \beta_3 & = & 0.2406\;(0.0601,0.4211)\\
               \hat \beta_4 & = & 0.0733\;(0.0402,0.1064) \\
              \end{array}
         \end{equation}
which, as expected, are very close to the true values used for simulation purposes.
     \item Additionally, we fit the instrumental sample, i.e.
     ${\bfg z}^{\rm max}_1 = {\bfg x}^{\rm max}_1+{\bfg y}_1$ to a GEV distribution, getting the following
        parameter estimates and 95\% confidence bounds:
         \begin{equation}\label{fitGEV3}
             \begin{array}{rcl}
               \hat \mu_1 & = & 5.4673 \;(5.4651,5.4695) \\
               \hat \psi_1 & = & 0.0060 \;(0.0045,0.0075) \\
               \hat \xi_1 & = & -0.1030\;(-0.1042 ,-0.1019). \\
              \end{array}
         \end{equation}

         Note that we fitted a GEV model because the shape parameter
         of the Pareto distribution fit is lower than $-1/2$.

    \item Using the three fitted models, the quantiles and their 95\% confidence intervals related to i) the reanalysis ($X$) data fit,
     ii) instrumental ($Z$) data fit using the proposed model, and iii) instrumental ($Z$) data fit using the GEV fit, are calculated.
      \end{enumerate}

      Results for this case are shown in Figure~\ref{QuantilesPOT}, where the following observations are pertinent:
      \begin{enumerate}
        \item The Pareto-Poisson and the proposed models provide very good fits with
        respect to simulated data, which was an expected result due to
        the length of the data samples. However, the GEV fit to instrumental maxima is
         not appropriate for return periods longer than 20 years.
        \item The proposed mixture model contains all empirical return period values within their
        95\% confidence bands. This proves the consistency of the proposed model.
        \item In terms of confidence intervals, the proposed model provides slightly
        better results with respect to the confidence intervals of
        the GEV model.
      \end{enumerate}

\begin{figure}
\begin{center}
\includegraphics*[width=0.95\textwidth]{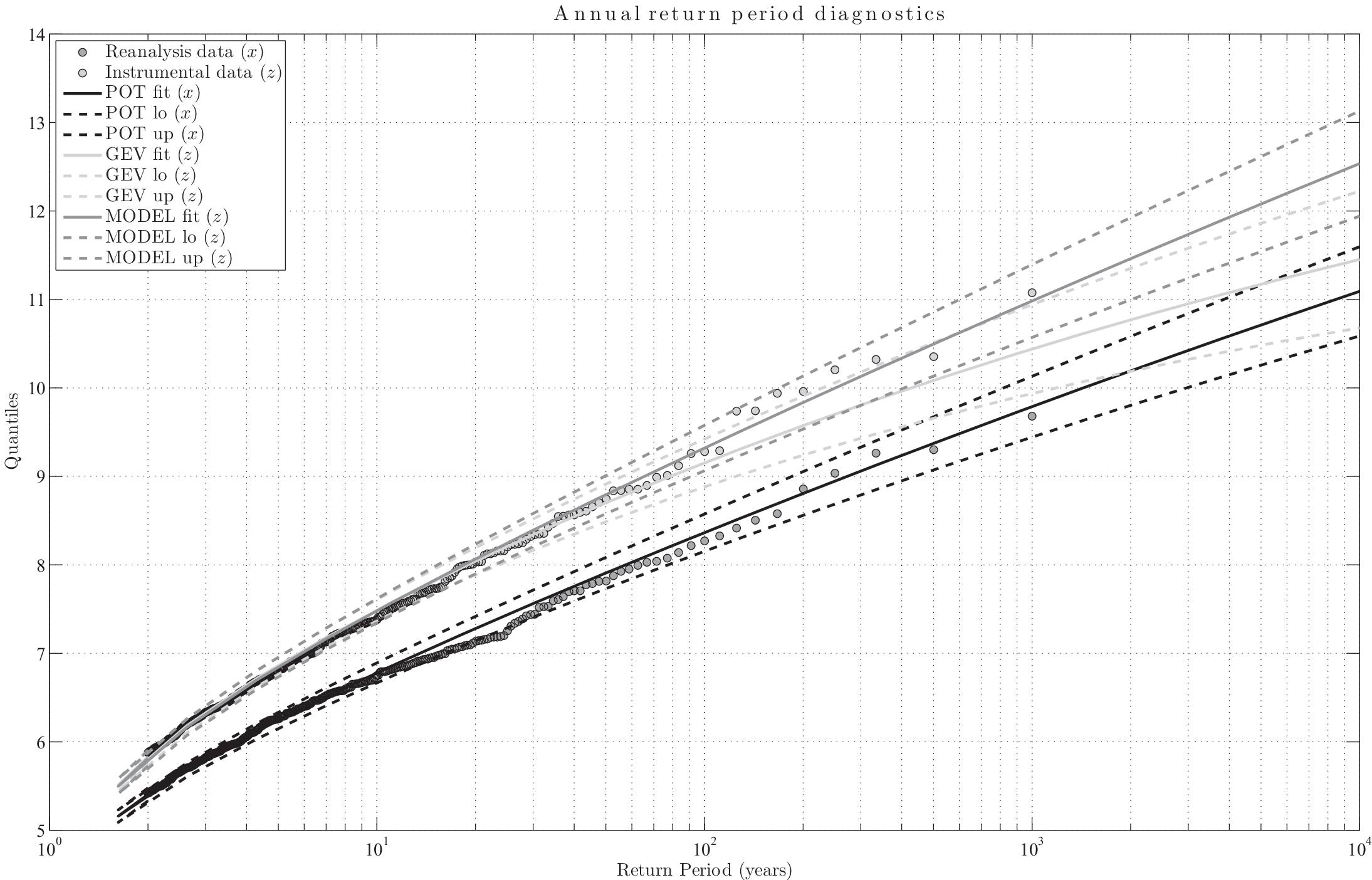}
\caption{\label{QuantilesPOT}Simulated data, quantiles and their 95\% confidence intervals related to i) the reanalysis ($X$) data fit,
     ii) instrumental ($Z$) data fit using the proposed model, and iii) instrumental ($Z$) data fit using the GEV fit.}
\end{center}
\end{figure}

\end{description}

Simulation results for both cases demonstrate the adequate functioning of the proposed model,
which will be further tested using real data.

\section{Realistic Illustrative Example}\label{s6}
In order to show the functioning of the proposed methodology in a realistic case study,
we have selected an specific location close to Bilbao Harbor (Northern coast of Spain).
 At this site, we have at our disposal i) hourly reanalysis significant wave height records from February 1, 1948 up to
January 1, 2011, and ii) buoy instrumental records from February 21, 1985 to July 13, 2009. Both
records are shown in Figure~\ref{f1}.

\subsection{Bilbao site using GEV}
In this subsection we analyze in detail the Bilbao record using as EV model the GEV.
Let consider the vectors ${\bfg x}$, ${\bfg x}^{\rm max}$, ${\bfg z}^{\rm max}$, and
${\bfg y}$ to be the reanalysis significant wave height records, the corresponding annual maxima,
the instrumental annual maxima, and the differences between ${\bfg x}^{\rm max}$ and ${\bfg z}^{\rm max}$
for those years where we have both records.
Using this information, we proceed to perform the following steps:
\begin{description}
  \item[Step 1:] Using the sample set $({\bfg x}^{\rm max}$, we fit the GEV distribution using the maximum
  likelihood method, i.e. by maximizing the log-likelihood function. The following parameter estimates
   and 95\% confidence bounds are obtained:
\end{description}
\begin{equation}\label{fitGEV4}
   \begin{array}{rcl}
     \hat \mu_x & = & 5.1046\;(4.9497,5.2596) \\
     \hat \psi_x & = &   -0.5173\;(-0.7125,-0.3222) \\
     \hat \xi_x & = & 0. \\
    \end{array}
\end{equation}

\begin{figure}[ht]
\begin{center}
\includegraphics*[width=0.95\textwidth]{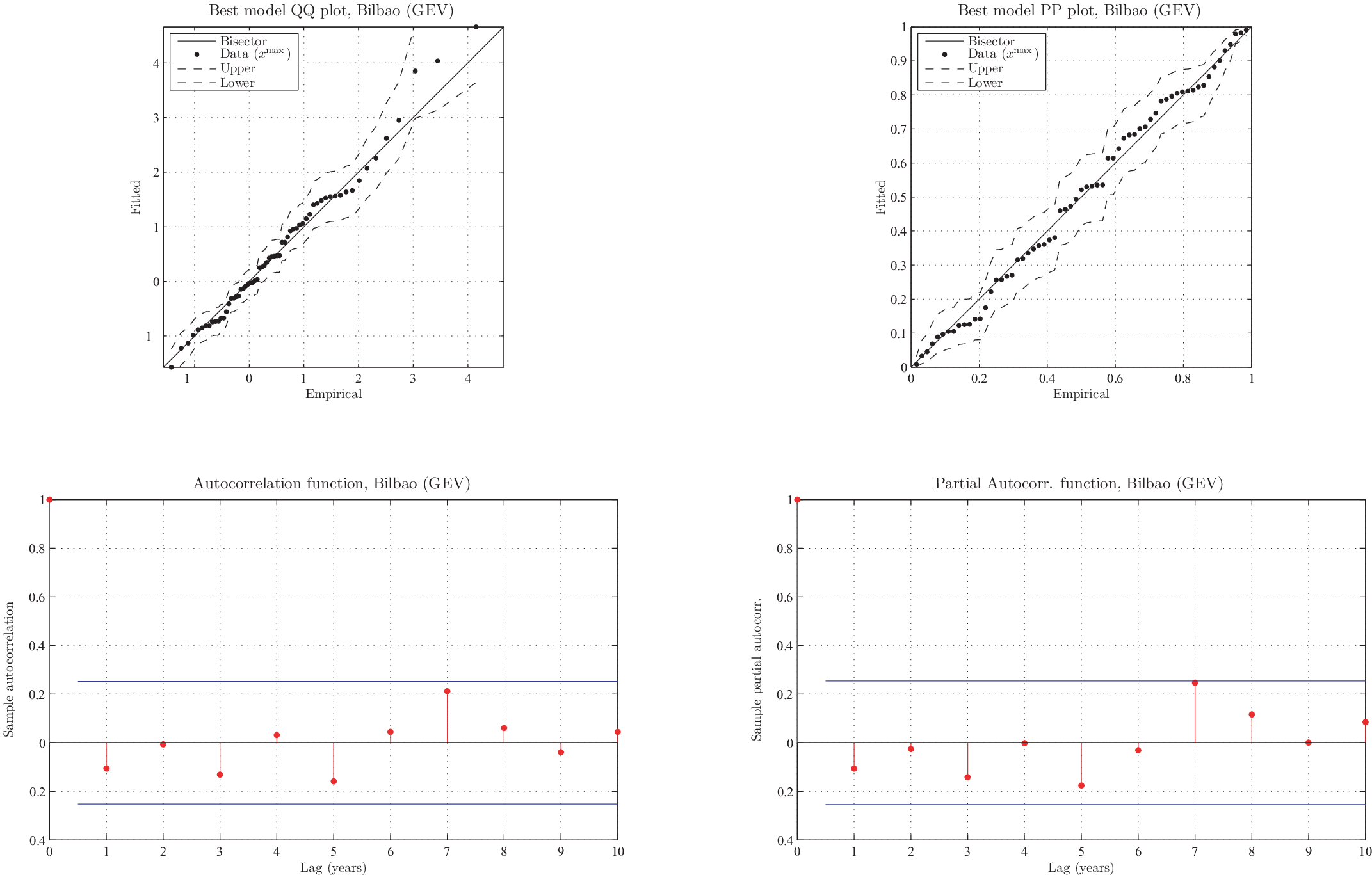}
\caption{\label{XmaxFitdiagBilbao} Diagnostic plots for the GEV fit related
 Bilbao site reanalysis maxima (${\bfg x}^{\rm max}$): i) PP plot, ii) QQ plot, iii) autocorrelation function,
 and iv) partial autocorrelation function.}
\end{center}
\end{figure}

This fit corresponds to the Gumbel case ($\hat \xi_x=0$). In Figure~\ref{XmaxFitdiagBilbao}
several diagnostic plots of the fitting are shown. Note that PP and QQ plots (panels above)
present good diagnostic statistics with points close to the diagonal. In addition, we
apply the one-sample Kolmogorov-Smirnov test with $0.05$ significance level for the transformed sample
${\bfg x}^{{\rm N}} = \Phi^{-1}\left[\hat F_{X^{}}({\bfg x}_{\rm max})\right]$.
Note that the $p$-value obtained is $0.9410$, so that the null hypothesis that the transformed
sample follow a standard normal distribution is accepted. This implies that the Gumbel fit is appropriate.
In addition, panels below of Figure~\ref{XmaxFitdiagBilbao} show, respectively, the
autocorrelation and partial autocorrelation functions of the transformed values.
Note that in both cases the autocorrelation and partial autocorrelation
functions for different time lags are within or close the confidence bands,
confirming that the values are uncorrelated. Finally, the Ljung-Box lack-of-fit hypothesis
test considering the null
hypothesis that no serial correlation at the lags 1, 2, 3, 4, and 5 years exist has been
 applied on the
     ${\bfg x}^{{\rm N}}$ sample. The $p$-values obtained for a 5\% significance level are
     ($0.3896$, $0.6897$,
$0.5897$, $0.7386$, $ 0.5840$), respectively. Note that since
 in all cases the $p$-values are higher than the significance level $0.05$,
the null hypothesis is accepted, which confirms the independence assumption between reanalysis annual maxima.

\begin{description}
  \item[Step 2:] Using the samples $({\bfg x}^{\rm max},{\bfg y})$ and assuming model
  (\ref{model1}) for the conditional mean and standard deviations, we fit the
   regression model by maximizing the log-likelihood function (\ref{Rmaxlikel}), getting the following
        parameter estimates and 95\% confidence bounds:
\end{description}
\begin{equation}\label{fitHET3}
   \begin{array}{rcl}
               \hat \beta_1 & = & -0.0219\;(-1.8532,1.8094)\\
               \hat \beta_2 & = & 0.1111\;(-0.2482,0.4705) \\
               \hat \beta_3 & = & -0.9966\;(-2.1516, 0.1585)\\
               \hat \beta_4 & = & 0.2894 \;(0.0608,0.5179) \\
    \end{array}
\end{equation}

\begin{figure}
\begin{center}
\includegraphics*[width=0.95\textwidth]{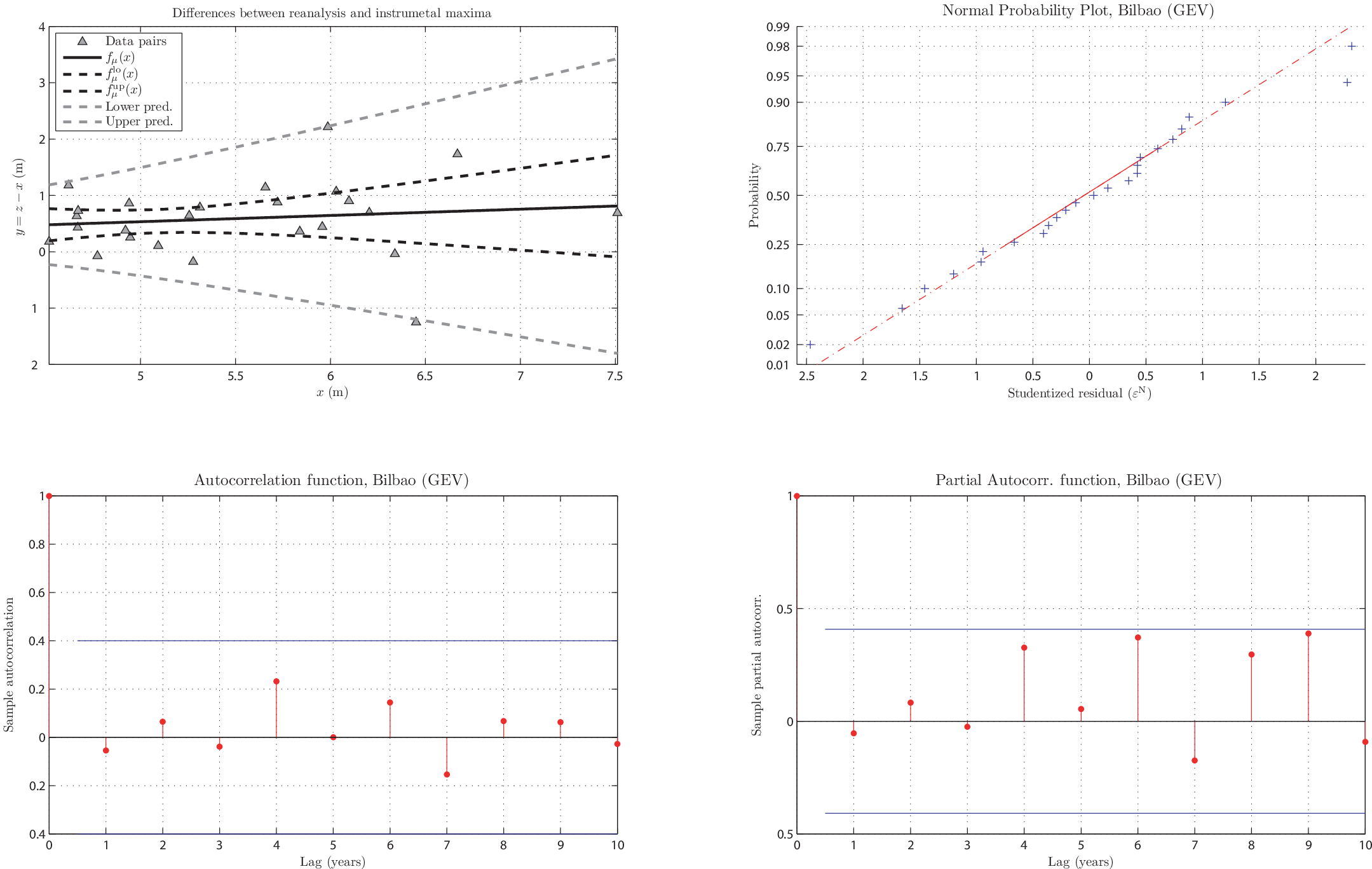}
\caption{\label{RegressionDiagnosBilbao} Diagnostic plots for the regression fit related
 Bilbao site $({\bfg x}^{\rm max},{\bfg y})$: i) data pairs, mean values, upper and lower bounds for both
 expected values and predicted response, ii) normal probability plot of studentized residuals,
  iii) autocorrelation function of studentized residuals ,
 and iv) partial autocorrelation function of studentized residuals.}
\end{center}
\end{figure}

Figure~\ref{RegressionDiagnosBilbao} shows different diagnostic plots for the regression model fitted.
Upper left panel presents the scatter plot (triangle dots), the conditional mean response (black line), 95\% confidence bands
for the mean response (dashed black line), and 95\% confidence bands for the predicted values (dashed gray line).
To check the normality assumption for studentized residuals given by (\ref{normres}), upper right panel
shows the studentized residuals on a normal probability plot. Note that data points are aligned
with the normal fit, i.e they follow a standard normal distribution. To further reinforce this
statement, we perform the one-sample Kolmogorov-Smirnov test with $0.05$ significance level
for the studentized residuals, obtaining a $p$-value equal to $0.9967$, i.e. the sample
comes from a standard normal distribution.
Finally, panels below of Figure~\ref{RegressionDiagnosBilbao} show, respectively, the
autocorrelation and partial autocorrelation functions of the studentized residuals.
Note that in both cases the autocorrelation and partial autocorrelation
functions for different time lags are within the confidence bands,
confirming that the values are uncorrelated. This is reinforced by performing
the Ljung-Box lack-of-fit hypothesis
test at the lags 1, 2, 3, 4, and 5 years.
The $p$-values obtained for a 5\% significance level are
     ($0.7730$, $0.9016$,
$0.9686$, $0.7379$, $ 0.8508$), respectively, and the independence hypothesis between
data is accepted.

\begin{description}
  \item[Step 3:] Finally, for comparison purposes we fit the sample ${\bfg z}^{\rm max}$
  to the GEV distribution. The following parameter estimates
   and 95\% confidence bounds are obtained:
\end{description}
\begin{equation}\label{fitGEV5}
   \begin{array}{rcl}
     \hat \mu_z & = & 5.6301\;(5.2956 , 5.9646) \\
     \hat \psi_z & = & -0.2090\;(-0.52745,0.1094) \\
     \hat \xi_z & = & 0. \\
    \end{array}
\end{equation}

\begin{figure}
\begin{center}
\includegraphics*[width=0.95\textwidth]{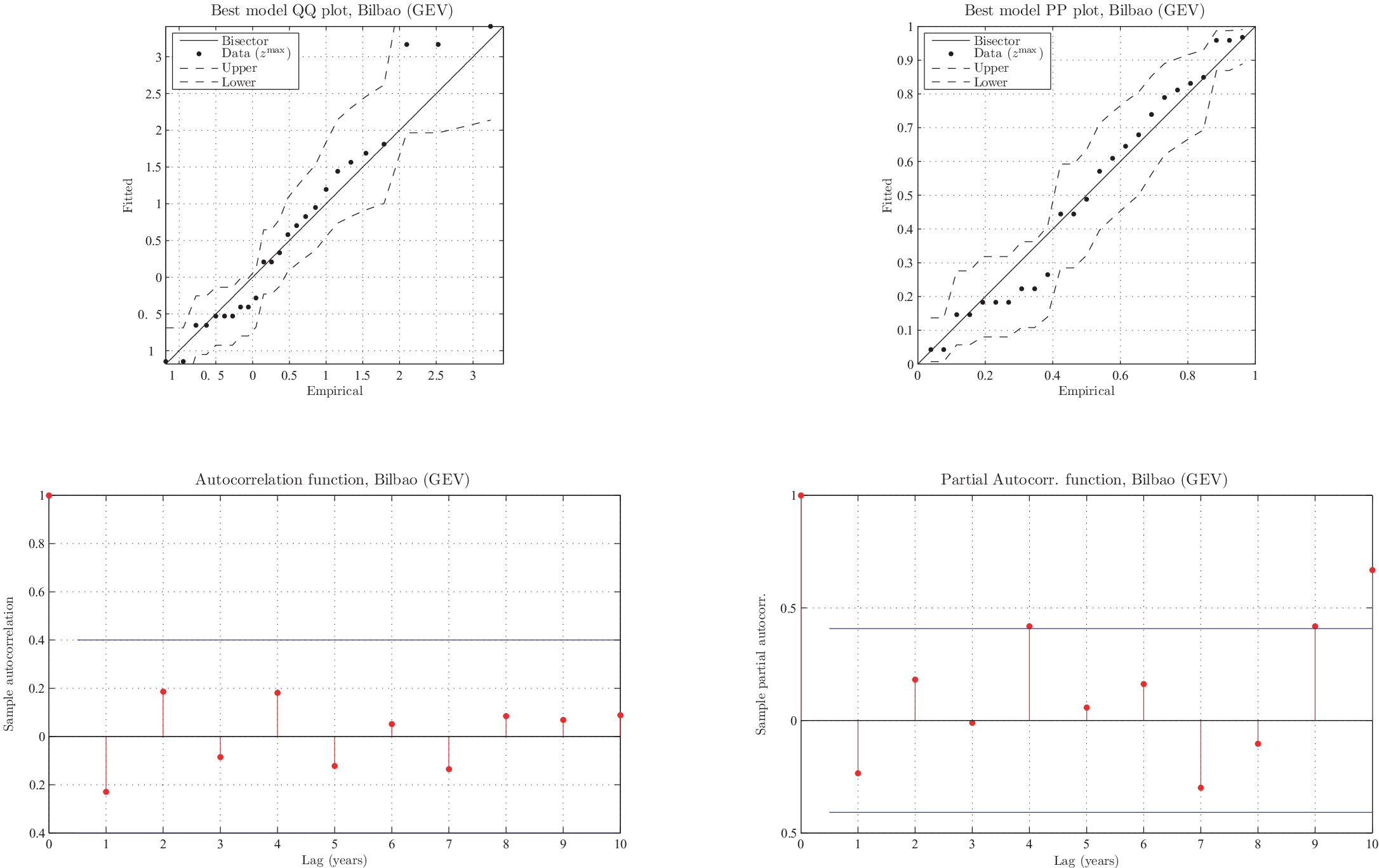}
\caption{\label{ZmaxFitdiagBilbao} Diagnostic plots for the GEV fit related
 Bilbao site instrumental maxima (${\bfg x}^{\rm max}$): i) PP plot, ii) QQ plot, iii) autocorrelation function,
 and iv) partial autocorrelation function.}
\end{center}
\end{figure}

The fit also corresponds to the Gumbel case ($\hat \xi_z=0$). In Figure~\ref{ZmaxFitdiagBilbao}, analogous
to the reanalysis maxima fit,
several diagnostic plots of the fitting are shown.
The fit is considered appropriate because, the one-sample Kolmogorov-Smirnov test with $0.05$
significance level for the transformed sample
${\bfg z}^{{\rm N}} = \Phi^{-1}\left[\hat F_{Z^{}}({\bfg z}_{\rm max})\right]$, allows accepting the null
hypothesis. The associated $p$-value is $0.6830$.
Analogously, the
autocorrelation and partial autocorrelation functions of the transformed values, shown
in panels below of Figure~\ref{ZmaxFitdiagBilbao}, indicate that the values
 are reasonable uncorrelated. This result is confirmed using the Ljung-Box lack-of-fit hypothesis
test considering the null
hypothesis that no serial correlation at the lags 1, 2, 3, 4, and 5 years exist.
 The $p$-values obtained for a 5\% significance level are
     ($0.2243$, $0.2877$,
$0.4378 $, $0.4377$, $ 0.5101$), respectively, confirming the independence assumption
between instrumental annual maxima.

\begin{description}
  \item[Step 4:] Using the information given by the three model fitted on previous steps, we calculate the
  return period values using: i) reanalysis maxima information, ii) instrumental maxima information,
  iii) reanalysis and instrumental maxima through the method proposed in this paper.
\end{description}

\begin{figure}[ht]
\begin{center}
\includegraphics*[width=0.95\textwidth]{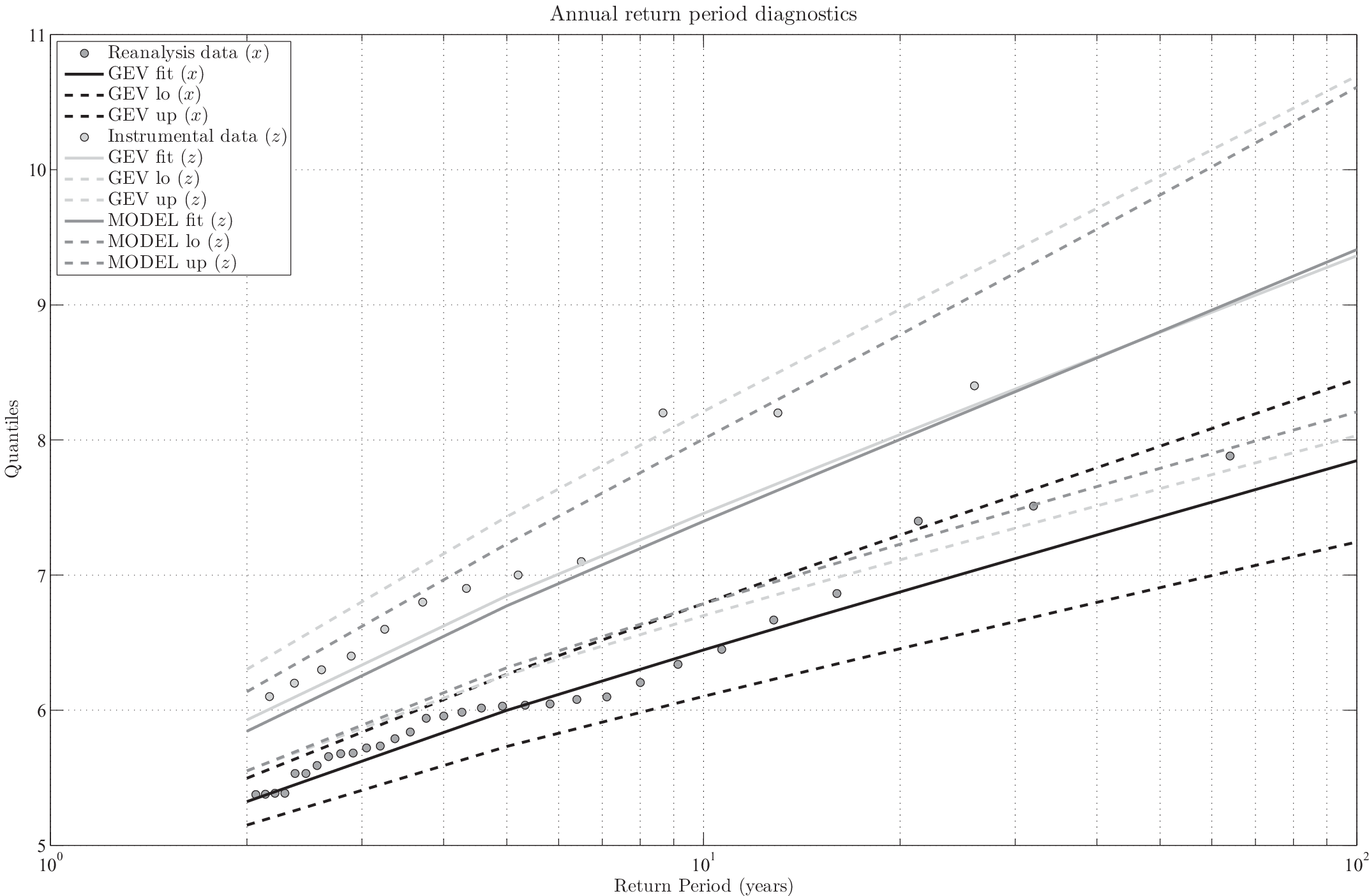}
\caption{\label{QuantilesBilbaoGEV} Annual return period values from: i) reanalysis data,
ii) instrumental data, iii) reanalysis GEV fitted model, iv) instrumental GEV fitted model,
and v) reanalysis and instrumental fitted model. For the models 95\% confidence bands are also
 plotted.}
\end{center}
\end{figure}

Results are summarized in Figure~\ref{QuantilesBilbaoGEV}, where the annual return periods from the models
and the data are shown. From this figure, the following comments are pertinent:
\begin{enumerate}
  \item The reanalysis fit (GEV($x$), black line) presents good agreement with respect to data, and the confidence bands
  are the narrowest from the three models. This result is obvious since the number of data values used for the
  fitting is the highest.

  \item Making extreme value analysis using reanalysis data lead to under predictions
  of return period values of about a meter, which is not acceptable from the engineering design perspective.

  \item Both the instrumental fit (GEV($z$)) and the proposed model fit (MODEL($z$)), are very close to
  each other, but data seems to present a bias. However, most of the data are within the
   confidence bands. Note that both models present the same return period value for 50 years.

   \item Confidence bands for the proposed model (MODEL($z$)) are always narrower than
   those for the instrumental fit (GEV($z$)), and are included between them.
   This proves that the proposed method decreases
    the uncertainty on return period values predictions.
\end{enumerate}

\section{Conclusions}\label{s7}
The model proposed in this paper allows to make extreme value analysis merging together reanalysis
and instrumental data. The proposed method has the following characteristics:
\begin{enumerate}
  \item The model is supported by probability distribution and non-linear regression theory.

  \item The hypothesis required to consider the proposed mixed model to be valid for EV analysis
  are properly established. In addition,
  several diagnostic plots and hypothesis tests are proposed to check whether these assumptions are held by the data.

  \item Numerical cases prove that the use of the proposed procedure provides more accurate return period values
  reducing their uncertainty.

  \item The model is very flexible, not only in terms of the marginal EV probability density function
  selected for the study of reanalysis maxima, but also for the regression model, which may deal with homoscedastic,
  heteroscedastic, linear and nonlinear models.
\end{enumerate}

\begin{acknowledgements}
This work was partly funded by projects ``AMVAR'' (CTM2010-15009) from Spanish Ministry MICINN,
by project C3E (200800050084091) from the Spanish Ministry MAMRM
 and by project MARUCA (E17/08) and OCTOPOS from the Spanish Ministry MF.
 R. M\'{\i}nguez is also indebted to the Spanish Ministry MICINN for the funding provided within
 the ``Ramon y Cajal'' program.
The authors acknowledge to Puertos del Estado the availability REDCOS coastal buoy
 network for this study.
\end{acknowledgements}

{\appendix

\section{Estimated parameters confidence intervals}\label{apenA}
The estimates $\hat{\bfg \theta}_X$ and $\hat{\bfg \beta}$ that maximize the log-likelihood functions
(\ref{fLOGlik}) or (\ref{fLOGlikParPoi}) and (\ref{Rmaxlikel}), respectively,
can be obtained using any of the available solvers for nonlinear programming. For specific details about the
heteroscedastic model and algorithms to solve (\ref{Rmaxlikel}) see \cite{MinguezRLM:11}.

The estimated parameters $\hat{\bfg \theta}_X$ and $\hat{\bfg \beta}$ correspond to mean values,
and assuming that observational errors are normally distributed, the estimated parameter vectors are distributed
as follows:
\begin{equation}\label{varcovCPar}
   {\bfg \theta}_X \sim N\left(\hat {\bfg \theta}_X,\Sigma_{{\bfg\theta}_X}\right);\quad  {\bfg \beta} \sim N\left(\hat {\bfg \beta},\Sigma_{\bfg\beta}\right),
\end{equation}
where $N$ denotes the multivariate normal distribution, and $\Sigma_{{\bfg\theta}_X}$ and $\Sigma_{\bfg\beta}$ are
 the variance-covariance matrix of the parameter estimates. Using the method of maximum likelihood,
if $\ell(\cdot)$ is twice differentiable with respect to estimated parameters, and under certain regularity conditions
which are often satisfied in practice (\cite{LehmannC:98}).
The parameters covariance matrix is equal to the inverse of the \emph{Fisher information matrix}
(${\bfg I}_{{\bfg \theta}_X},{\bfg I}_{\bfg \beta}$),
which is equal to the Hessian matrix of the log-likelihood function with the sign changed:
\begin{equation}\label{FisherIM}
 {\bfg I}_{{\bfg \theta}_X}= -\fraca{\partial^2 \ell({\bfg \theta}_X;{\bfg x}_{\rm max})}{\partial^2 {\bfg \theta}_X};\quad
  {\bfg I}_{\bfg \beta}= -\fraca{\partial^2 \ell(\bfg \beta;{\bfg x},{\bfg y})}{\partial^2 \bfg \beta}.
\end{equation}

The $(1-\alpha)$ confidence interval for each parameter is equal to:
\begin{equation}\label{condinte}
\begin{array}{ccc}
   \theta_{X_j}^{\rm up}&=&\hat\theta_{X_j}+ t_{(1-\alpha/2,n_x-n_{p_x}-1)}\hat\sigma_{X_j},\;j=0,1,\ldots,n_{p_x} \\
 \theta_{X_j}^{\rm lo}&=&\hat\theta_{X_j}- t_{(1-\alpha/2,n_x-n_{p_x}-1)}\hat\sigma_{X_j},\;j=0,1,\ldots,n_{p_x},  \\
  \beta_j^{\rm up}&=&\hat\beta_j+ t_{(1-\alpha/2,n_y-n_p-1)}\hat\sigma_j,\;j=0,1,\ldots,n_p \\
 \beta_j^{\rm lo}&=&\hat\beta_j- t_{(1-\alpha/2,n_y-n_p-1)}\hat\sigma_j,\;j=0,1,\ldots,n_p ,
\end{array}
\end{equation}
where $t_{(1-\alpha/2,n_{df})}$ is the Student's $t$-distribution $(1-\alpha/2)$ quantile with $n_{df}$
degrees of freedom and $\hat\sigma_{X_j}$ and $\hat\sigma_j$ are the
corresponding estimated standard deviations for parameters $j$ (square root of the corresponding diagonal term
 in $\bfg \Sigma_{{\bfg\theta}_X}$ and $\bfg \Sigma_{{\bfg\beta}}$, respectively).

\section{Quantile confidence intervals}\label{apenB}
From the engineering design perspective, it is of great interest the calculation of return period values for
different time spans ($T_d$) (usually in years), which correspond, within the extreme value model selected, to quantiles associated with
the following probability of not exceedance $q_{T_d} = 1-1/T_d$.

For the reanalysis case, these estimated quantiles $\hat x_q$ are calculated
depending on the EV analysis model selected. For the proposed model, combining instrumental and reanalysis information,
quantiles are obtained solving the implicit equation (\ref{quan}).

If we are interested in calculating the
confidence bands for reanalysis quantiles $x_q$, it is known that for large sample sizes $n_{x}$,
the quantile $x_q$ is asymptotically normal, and thus, the delta method can be applied as follows:
\begin{equation}\label{deltmeth}
     x_q \sim N(\hat x_q,\nabla_{{\bfg \theta}_X}^T x_q\bfg \Sigma_{{\bfg\theta}_X}\nabla_{{\bfg \theta}_X} x_q),
\end{equation}
where $\nabla_{{\bfg \theta}_X}x_q$ is the $n_{p_x}$ vector of partial derivatives of quantile
expressions with respect to ${\bfg \theta}_X$.

Note that equation (\ref{deltmeth}) allows obtaining the estimated variance $\hat\sigma^2_{x_q}$ of the quantile, and
the confidence intervals become:
\begin{equation}\label{condintexqX}
\begin{array}{ccc}
   x_q^{\rm up}&=&\hat x_q+ t_{(1-\alpha/2,n_x-n_{p_x}-1)}\hat\sigma_{x_q}, \\
 x_q^{\rm lo}&=&\hat x_q- t_{(1-\alpha/2,n_x-{p_x}-1)}\hat\sigma_{x_q},
\end{array}
\end{equation}

For the proposed model, the process is analogous, we use the delta method in order to get the estimated variance
of the corresponding quantile $z_q$:
\begin{equation}\label{deltmethZ}
     z_q \sim N(\hat z_q,\nabla_{\left({\bfg \theta}_X;{\bfg \beta}\right)}^T z_q\bfg \Sigma_{\left({\bfg\theta}_X; {\bfg \beta}\right)}
     \nabla_{\left({\bfg \theta}_X;{\bfg \beta}\right)} z_q,
\end{equation}
where $\nabla_{\left({\bfg \theta}_X;{\bfg \beta}\right)}z_q$ is the $n_{p_x}+n_{p}$ vector of partial derivatives
 of quantiles from solving the implicit equation (\ref{quan}) with respect to ${\bfg \theta}_X$ and
 ${\bfg \beta}$. $\Sigma_{\left({\bfg\theta}_X; {\bfg \beta}\right)}$ is the variance-covariance matrix of all the estimated
 parameters, including the extreme value and the regression models. Since both models are independent by definition,
 it is equal to:
 \begin{equation}\label{MatVarModelZ}
     \Sigma_{\left({\bfg\theta}_X; {\bfg \beta}\right)} =
     \left(\begin{array}{cc}
       \bfg \Sigma_{{\bfg\theta}_X} & {\bfg 0} \\
       {\bfg 0}  & \bfg \Sigma_{{\bfg\beta}}
     \end{array}\right).
 \end{equation}

Note that the required derivatives for the reanalysis case are easily obtained analytically, however, for the
 composed model is a challenge in itself. For this reason, these are obtained numerically by finite differences:
\begin{equation}\label{numder}
   \fraca{\partial z_q}{\partial \gamma}=\fraca{z_q(\gamma(1+\epsilon))-z_q(\gamma(1-\epsilon))}{\epsilon \gamma},
\end{equation}
where $\gamma$ represents the corresponding parameter and $\epsilon=10^{-6}$.

}


\end{document}